%% file: main.tex
\documentclass[sigconf]{acmart}

\AtBeginDocument{%
    \providecommand\BibTeX{{%
        \normalfont B\kern-0.5em{\scshape i\kern-0.25em b}\kern-0.8em\TeX}
    }
}

\usepackage{ifthen}

\usepackage[capitalize]{cleveref}
\usepackage[binary-units]{siunitx}
\usepackage{natbib}
\usepackage{tikz}
\usetikzlibrary{calc}
\usetikzlibrary{intersections}

\usepackage[disable]{todonotes}

\newcommand{\todosolved}[1]{\todo[inline, color=green]{#1}}

\usepackage[font=footnotesize,labelfont=bf]{caption}
\usepackage{glossaries}

\newif\ifarxiv
\arxivtrue

\input{glossary.tex}

\acmConference[NICE '20]{Neuro Inspired Computational Elements Workshop}{March 17--20, 2020}{Heidelberg, Germany}
\acmBooktitle{NICE '20: Neuro-Inspired Computational Elements Workshop, March 17--20, 2020, Heidelberg, Germany}

\title{Closed-loop experiments on the BrainScaleS-2 architecture}

\author{
    K.~Schreiber,$\ $
    T.~C.~Wunderlich,$\ $
    C.~Pehle
}
\affiliation{%
    \footnotesize
    \institution{Kirchhoff-Institute for Physics, Heidelberg University}
    \city{Heidelberg}
    \state{Germany}
    \postcode{69120}
}

\author{
    M. A. Petrovici
}
\affiliation{%
    \footnotesize
    \institution{Department of Physiology, University of Bern}
    \city{Bern}
    \state{Switzerland}
    \postcode{3012}
}
\author{
    J.~Schemmel,$\ $
    K. Meier
}
\affiliation{%
    \footnotesize
    \institution{Kirchhoff-Institute for Physics, Heidelberg University}
    \city{Heidelberg}
    \state{Germany}
    \postcode{69120}
}

\renewcommand{\shortauthors}{K.~Schreiber, et al.}

\date{\today}

\setcopyright{rightsretained}

\begin{document}

\begin{abstract}
    \input{content/abstract}
\end{abstract}

\copyrightyear{2020} 
\acmYear{2020} 
\acmConference[NICE '20]{Neuro-inspired Computational ElementsWorkshop}{March 17--20, 2020}{Heidelberg, Germany}
\acmBooktitle{Neuro-inspired Computational Elements Workshop (NICE
'20), March 17--20, 2020, Heidelberg,
Germany}\acmDOI{10.1145/3381755.3381776}
\acmISBN{978-1-4503-7718-8/20/03}

\ccsdesc[500]{Hardware~Analog and mixed-signal circuits}

\keywords{Closed-loop, neuromorphic hardware, path integration, reinforcement learning, neurorobotics}

\maketitle

\section{Introduction}
\label{sec:intro}
\input{content/introduction}
\section{Reinforcement learning}
\label{sec:pong}
\input{content/pong}
\section{Accelerated robotics}
\label{sec:playpen}
\input{content/playpen}
\section{Insect navigation}
\label{sec:insects}
\input{content/insects}
\section{Conclusion}
\label{sec:discussion}
\input{content/discussion}

\begin{acks}
\input{content/ack}%
\end{acks}

\ifarxiv
    \input{main.bbl}

\else
    \bibliography{bib}
\fi

\end{document}

%% file: glossary.tex
\newacronym{ppu}{PPU}{plasticity processing unit}
\newacronym{padi}{PADI}{parallel driver interface}

\newacronym{sram}{SRAM}{static random-access memory}
\newacronym{adc}{ADC}{analog-to-digital converter}
\newacronym{cadc}{CADC}{column-parallel analog-to-digital converter}
\newacronym{dac}{DAC}{digital-to-analog converter}
\newacronym{dut}{DUT}{design under testing}
\newacronym{asic}{ASIC}{application-specific integrated circuit}
\newacronym{fpga}{FPGA}{field-programmable gate array}
\newacronym{vlsi}{VLSI}{very-large-scale integration}
\newacronym{simd}{SIMD}{single instruction, multiple data}
\newacronym{ocp}{OCP}{Open Core Protocol}
\newacronym{soc}{SoC}{system on a chip}
\newacronym{sta}{STA}{static timing analysis}
\newacronym{gals}{GALS}{globally asynchronous locally synchronous}
\newacronym{dpi}{DPI}{SystemVerilog direct programming interface}
\newacronym{pll}{PLL}{phase-locked loop}
\newacronym{fifo}{FIFO}{First-In-First-Out Buffer}

\newacronym{psc}{PSC}{post-synaptic current}
\newacronym{psp}{PSP}{post-synaptic potential}
\newacronym{stp}{STP}{short-term plasticity}
\newacronym{stdp}{STDP}{spike-timing-dependent plasticity}
\newacronym{rstdp}{R-STDP}{reward-modulated spike-timing-dependent plasticity}
\newacronym{lif}{LIF}{leaky integrate-and-fire}
\newacronym{adex}{AdEx}{adaptive exponential leaky integrate-and-fire}

\newacronym{mc}{MC}{Monte Carlo}

%% file: content/abstract.tex
The evolution of biological brains has always been contingent on their embodiment within their respective environments, in which survival required appropriate navigation and manipulation skills.
Studying such interactions thus represents an important aspect of computational neuroscience and, by extension, a topic of interest for neuromorphic engineering.
Here, we present three examples of embodiment on the BrainScaleS-2 architecture, in which dynamical timescales of both agents and environment are accelerated by several orders of magnitude with respect to their biological archetypes.

%% file: content/introduction.tex
Neuromorphic engineering aims at overcoming certain limitations of traditional computer architectures by reproducing particular aspects of structure and function of biological neural networks in VLSI.
Within this context, it is important to note that all biological brains are part of a body that interacts with its environment in various ways: information continuously flows from a diverse range of sensory organs or cells into the nervous system.
The nervous system processes this information and in turn provides signals to organs or cells that are involved in motion or communication, resulting in actions within the environment.
This interplay of information exchange and processing appears to be inseparably linked to the working principles of biological brains and forms a closed loop of environment, body, and brain.
These important aspects of so-called embodied cognition have received increasing attention in cognitive science and related fields\cite{wilson2002six}, evidently pertaining to both biological and artificial brains.
For versions of the latter running on analog neuromorphic hardware, embodiment carries particular constraints, as the continuous-time dynamics of physical analog neurons and synapses cannot be accelerated, slowed-down or even halted as easily as for their digital counterparts.

Here, we discuss three different implementations of closed-loop experiments on the BrainScaleS-2 system~\cite{aamir2018lifarray, aamir2018adex, ISCAS2020billaudelle}: a neural agent playing the game of pong, a robotic application in which the neural network is connected to a pantographic robotic system, and an insectoid agent that performs a path integration task in a 2D environment.
By allowing continuous access to the spike I/O of the analog neural core, as well as digital online updates of network parameters, the hybrid architecture of BrainScaleS-2 lends itself particularly well to the study of neural agents that behave and learn within an interactive environment.

%% file: content/pong.tex
Reinforcement learning represents a natural fit for closed-loop experiments in which an agent tries to maximize a reward signal based on its actions within an environment.
It was recently shown that BrainScaleS-2 can be used to implement closed-loop reinforcement learning, where all components of the loop, including the virtual environment simulation, are computed on-chip, thus creating a fully autonomous setup \cite{wunderlich2019demonstrating}.

\begin{figure}
    \centering
    \includegraphics[width=0.48\textwidth]{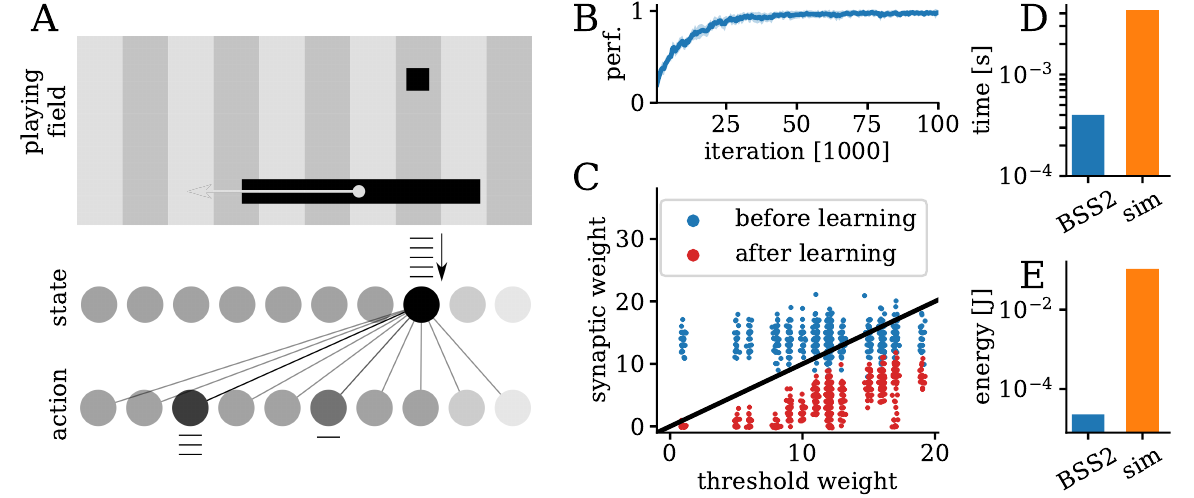}
    \caption{
        \textbf{Reinforcement learning with reward-modulated STDP.}
        \textbf{A)} The \gls{ppu} simulates a simplified version of Pong.
                    The horizontal position of the ball serves as input for a 2-layer neural network, with the resulting output dictating the target paddle position.
                    The network receives reward based on its aiming accuracy.
        \textbf{B)} Playing performance during learning.
        \textbf{C)} Synaptic depression automatically adapts to the excitability of neurons.
        \textbf{D, E)} Wall-clock duration and power consumption of a single iteration on BrainScaleS-2 (blue) and an equivalent software simulation using NEST (orange).
        }
    \label{fig:pong}
\end{figure}

Our referenced work uses a three-factor learning rule called R-STDP~\cite{fremaux2010functional,fremaux2016neuromodulated}.
This rule multiplies a scalar reward signal, akin to a global neuromodulator such as dopamine, with an STDP-like synaptic eligibility trace.
In our case, the latter is computed locally at each synapse and in an analog fashion \cite{friedmann2016hybridlearning}.

The virtual environment is a simplified version of the Pong video game, as shown in \cref{fig:pong}A.
The spiking neural network dynamics are emulated by the neuromorphic substrate while the embedded plasticity processor takes on a dual role:
it both simulates the game dynamics and computes synaptic weight updates using R-STDP, creating a fully autonomous setup.

During training, the network learns to keep the paddle approximately centered under the ball (\cref{fig:pong}B).
Implicitly, the experiment also demonstrates how learning can compensate fixed-pattern noise in the analog neuro-synaptic circuits (\cref{fig:pong}C): we found that while the excitability of uncalibrated neurons varied significantly due to mismatch effects, synapses that would negatively impact correct tracking of the ball were systematically depressed below the individual firing threshold of their postsynaptic neuron.
Furthermore, this setup demonstrates the speed and power advantages of the BrainScaleS-2 architecture compared to software simulations (\cref{fig:pong}D/E).

%% file: content/playpen.tex
\begin{figure}
    \definecolor{fg}{rgb}{0.0, 0.0, 0.0}
    \definecolor{bg}{rgb}{1.0, 1.0, 1.0}
    \definecolor{highlight}{HTML}{FB2CF9}
    \newcommand{\pgfextractangle}[3]{%
    \pgfmathanglebetweenpoints{\pgfpointanchor{#2}{center}}
                            {\pgfpointanchor{#3}{center}}
    \global\let#1\pgfmathresult
    }
    \centering
    \begin{tikzpicture}[scale=0.6, every node/.style={transform shape}]
        \node (playpen legend) at (-3, 6) {\huge A};

        \newcommand*{\ALPHA}{50.0}%
        \newcommand*{\BETA}{10.0}%
        \newcommand*{\RS}{3.0}%
        \newcommand*{\R}{4.0}%

        \draw[fill=fg!20!bg] (1.3, 1.5) rectangle ++(5.8, 4.6);
        \node[draw, rectangle, minimum height=5cm, minimum width=6cm, anchor=south west] (screen) at (1.2, 1.2) {Screen};

        \draw (0, 0.2) rectangle ++(.8, 2);
        \node[draw, circle, radius=.1, fg!20!bg, fill=fg] (mx) at (.4, 1.2) {MX};
        \draw (1.2, 0) rectangle ++(2, .8);
        \node[draw, circle, radius=.1, fg!20!bg, fill=fg] (my) at (2.2, .4) {MY};

        \coordinate (jx) at ($ (mx) + (\ALPHA:\RS) $);
        \coordinate (jy) at ($ (my) + (\BETA:\RS) $);

        \path[overlay, name path=c1] (jx) circle (\R);
        \path[overlay, name path=c2] (jy) circle (\R);
        \path[name intersections={of=c1 and c2, by={a, b}}];

        \draw[line width=2pt] (mx) -- (jx);
        \draw[line width=2pt] (my) -- (jy);
        \draw[line width=2pt] (jx) -- (a);
        \draw[line width=2pt] (jy) -- (a);

        \node[draw, circle, radius=.4, fg, fill=fg] (JX) at (jx) {};
        \node[draw, circle, radius=.4, fg, fill=fg] (JY) at (jy) {};

        \node[draw, circle, radius=.4, fg!20!bg, fill=fg, align=center] (pickup) at (a) {S};

        \draw[<->] (mx) ++(160:.6) arc (160:20:.6);
        \draw[<->] (my) ++(80:.6) arc (80:-80:.6);

        \node (view description) at (6, .2) {Top View};

        \newcommand*{\dangle}{6.5}%
        \pgfextractangle{\phi}{jy}{a}
        \draw[->, line width=1pt, color=fg] ([shift=(\phi+\dangle:\R)]jy) arc (\phi+\dangle:\phi+20:\R);
        \draw[->, line width=1pt, color=fg] ([shift=(\phi-\dangle:\R)]jy) arc (\phi-\dangle:\phi-20:\R);

        \pgfextractangle{\phi}{jx}{a}
        \draw[->, line width=1pt, color=fg] ([shift=(\phi+\dangle:\R)]jx) arc (\phi+\dangle:\phi+20:\R);
        \draw[->, line width=1pt, color=fg] ([shift=(\phi-\dangle:\R)]jx) arc (\phi-\dangle:\phi-20:\R);

        \node[draw, rectangle, text width=2cm, align=center] (spikein) at (-1.5, 5.6) {Sensor signal\\$\downarrow$\\Spikes};
        \node[draw, rectangle, minimum height=1.9cm, text width=2cm] (hicann) at (-1.5, 3.4) {BrainScaleS-2};
        \node[draw, rectangle, text width=2cm, align=center] (spikeout) at (-1.5, 1.15) {Spikes\\$\downarrow$\\Motor signal};
        
        \draw[line width=1pt, ->, color=highlight] (pickup) to [out=150, in=0] (spikein);
        \draw[line width=1pt, ->, color=highlight] (spikein) to [out=270, in=90] (hicann);
        \draw[line width=1pt, ->, color=highlight] (hicann) to [out=270, in=90] (spikeout);
        \draw[line width=1pt, ->, color=highlight] (spikeout) to [out=340, in=180] (my);
        \draw[line width=1pt, ->, color=highlight] (spikeout) to [out=360, in=180] (mx);

        \node (playpen legend) at (8.5, 6) {\huge B};

        \coordinate (o0) at (9, 6.2);
        \coordinate (o1) at (9, 4.7);
        \coordinate (o2) at (9, 3.2);
        \coordinate (o3) at (9, 1.7);
        
        \draw[fill=fg] ($(o0) + (0, 0)$) rectangle ($(o0) + (1.5, -1.2)$);
        \draw[color=bg, fill=bg] ($(o0) + (.75, -.6)$) circle (.4);

        \draw[fill=fg] ($(o1) + (0, 0)$) rectangle ($(o1) + (1.5, -1.2)$);
        \filldraw[even odd rule, inner color=bg, outer color=fg] ($(o1) + (.75, -.6)$) circle (.6);

        \draw[fill=fg] ($(o2) + (0, 0)$) rectangle ($(o2) + (1.5, -1.2)$);
        \draw[color=bg, line width=3pt] plot[smooth cycle] coordinates {($(o2) + (.2, -.4)$) ($(o2) + (.6, -.2)$) ($(o2) + (1., -.4)$) ($(o2) + (1.3, -.5)$) ($(o2) + (.7, -1.1)$) ($(o2) + (.3, -.7)$)};
        \draw[-latex, color=highlight, thin, line width=.5] plot[smooth, tension=1] coordinates {($(o2) + (.6, -.2)$) ($(o2) + (1., -.4)$) ($(o2) + (1.3, -.5)$) ($(o2) + (.7, -1.1)$)};

        \draw[fill=fg] ($(o3) + (0, 0)$) rectangle ($(o3) + (1.5, -1.2)$);
        \draw[color=bg, fill=bg, opacity=.3] ($(o3) + (.5, -.45)$) circle (.1);
        \draw[color=bg, fill=bg, opacity=.5] ($(o3) + (.7, -.6)$) circle (.1);
        \draw[color=bg, fill=bg, opacity=.7] ($(o3) + (.9, -.7)$) circle (.1);
        \draw[color=bg, fill=bg, opacity=1.] ($(o3) + (1.1, -.75)$) circle (.1);
        \draw[-latex, color=highlight] plot[smooth, tension=1] coordinates {($(o3) + (.5, -.45)$) ($(o3) + (.7, -.6)$) ($(o3) + (.9, -.7)$) ($(o3) + (1.1, -.75)$)};
            
	  \end{tikzpicture}
    \caption{
        \textbf{PlayPen: a pantographic robotic system.}
        A) Arbitrary optical content of a screen is captured by a sensor S and translated into a continuous afferent stream of spikes.
        This input is, in turn, used by spiking networks emulated on the BrainScaleS-2 system to control the position of the sensor by emitting spike trains that drive the motors MX and MY.
        B) Example screen contents for four possible experimental tasks: stay within a potential well, follow a gradient, follow a trace, follow a dot.
        \todosolved{MAP: ALL of the tikz :D But seriously: maybe we can discuss the fig a bit and tweak it here and there? $\to$ KS: Move boxes to the left. Check for symmetry/alignment. Insert fig B.}
    }
    \label{fig:playpen}
\end{figure}
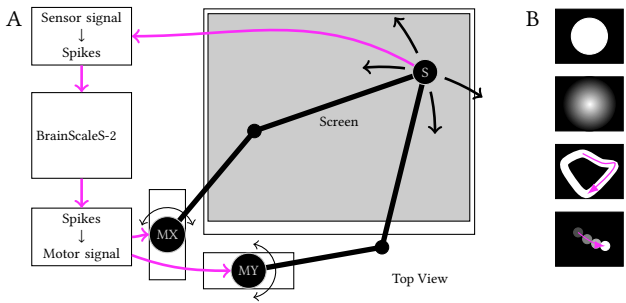

By implementing a fast low-latency link for peripheral spike input and output, we realized an experimental platform that provides BrainScaleS-2 with access to real-world robotic actuators and sensors.
Unlike common neurorobotics, the inherent thousandfold speed-up of BrainScaleS-2 demands exceptionally quick actuation and sensory feedback.
We therefore needed to reduce the weight -- and thus also the complexity -- of the mechanical components as much as possible.
The setup translates efferent spike trains into impulses that control two motors of a pantographic actuator that moves an optical sensor over a two-dimensional surface.
The data gathered by this sensor is in turn translated into an afferent spike train that is transmitted back to the neuromorphic core.
This simple construction, which we call the "PlayPen", provides an intuitive and easily configurable setup for the study of spike-based learning in the context of real-world robotic control.
\todosolved{MAP: I added ' which we call the "playPen" ' above and in the caption, because 1) I love the name and 2) I can reference it more easily in the discussion. $\to$ KS: ACK.}
The setup is depicted in \cref{fig:playpen} together with four example experiments.
\todosolved{MAP: Add 4 subpanels to panel B in Fig. 2 and a sentence here. $\to$ KS}

%% file: content/insects.tex
\begin{figure}
    \centering
    \includegraphics[width=0.48\textwidth]{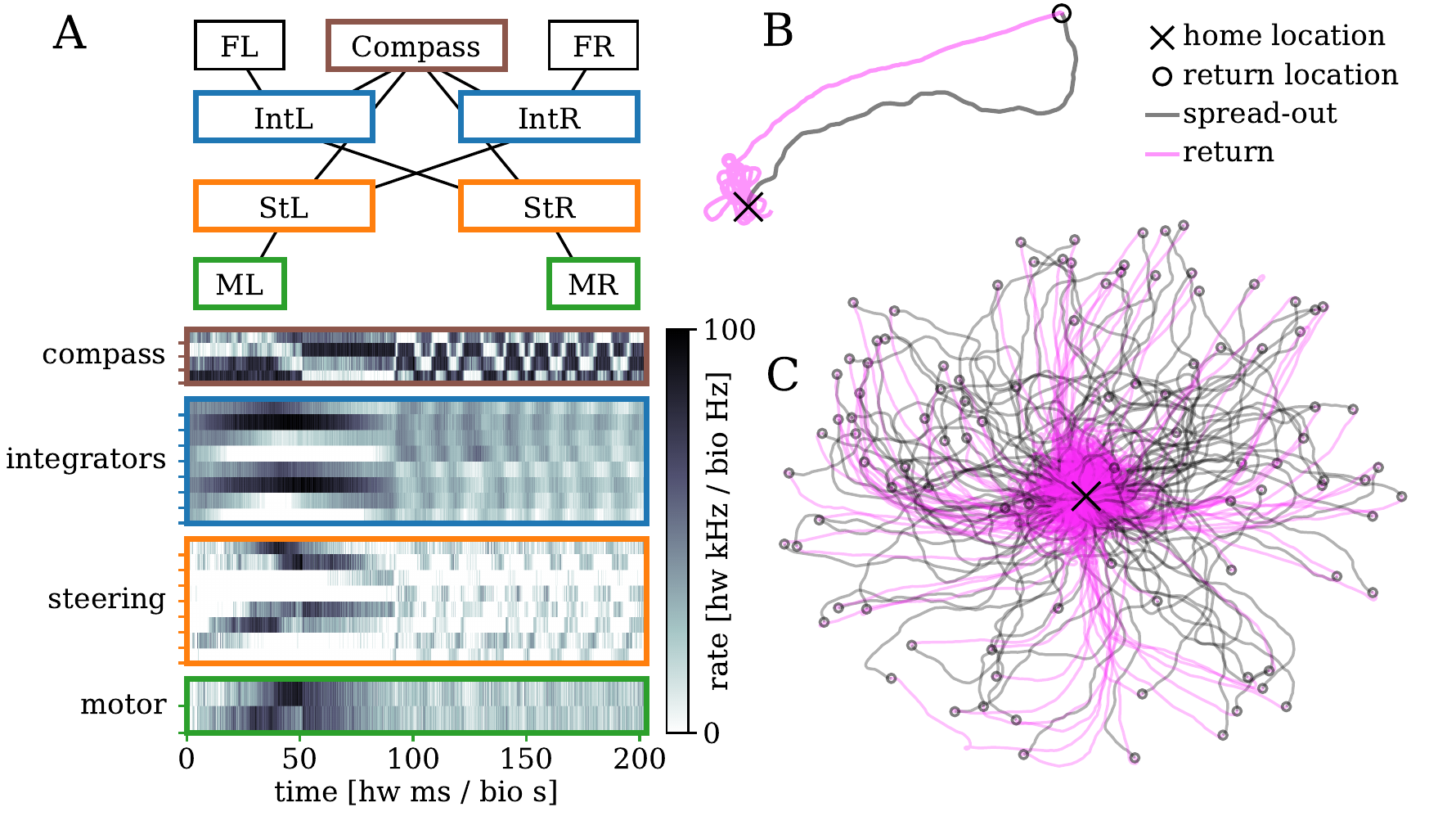}
    \caption{
        \textbf{Virtual insectoid agent performing path integration on BrainScaleS-2.}
        \textbf{A)} Network schematic and activity histogram.
                    The information flows from the sensory layer at the top through an integration and a steering layer to the motor neurons at the bottom.
                    R and L indicate the right and left side, respectively.
        \textbf{B)} A typical trajectory of the virtual insect which turns to random looping around the home position upon reaching it.
        \textbf{C)} Overlay of 100 trajectories like in B), each with a different random outbound journey.
    }
    \label{fig:insects}
\end{figure}

Recent developments in biological imaging have facilitated unprecedented insight into numerous functional aspects of insect brains \cite{chiang2011three, takemura2013visual, takemura2017connectome}, such as their navigational capabilities~\cite{neuser2008analysis}.
Based on physiological data from the bee's brain and following \cite{stone2017anatomically}, we emulated a network for path integration (\cref{fig:insects}A) that reproduces bees' ability to return to their nest's location after exploring the environment for sources of food.

At the beginning of each experiment, a virtual insect performed a random walk starting from a certain starting position.
During this phase, the modeled network had no effect on the insect's motion, but was provided with sensory data of the absolute head orientation and the optical flow field from the insect's eyes.
The insect's head orientation was encoded by four spike sources that each represented a cardinal direction similar to a compass.
The optical flow field was similarly represented by two spike generators that fired with a rate proportional to the optical flow as derived from the left and right eye (FL and FR).
In the second part of the experiment, the return phase, the insect's motion was determined by the two motor neurons in the network (ML and MR), which steered the insect by providing propulsion on the left or right hand side, similar to a tank drive.
Across multiple experiments, the emulated navigation network was able to reliably guide the insect back to its starting position (\cref{fig:insects}C).
As with Pong (Sec. 2), the plasticity processor was used to handle multiple tasks: the processing of synaptic modulations for the integrator neurons, the simulation of the environment, an emulation of all sensors including the corresponding spike stimuli, the translation ofneuronal data into actions of motion, and the entire experiment control.
The experiment can thus run entirely self-contained on the BrainScaleS-2 system.

%% file: content/discussion.tex
\todosolved{MAP: Should we call this section "Conclusion"? $\to$ KS: ACK.}

\todosolved{MAP: Suggestion:\\
    \vspace{5pt}
    We outlined three agent-environment interaction scenarios using spiking neural networks emulated on the BrainScaleS-2 architecture.
    In two of these, the sensors and actuators of an embodied neural agent, along with its environment, were also emulated on-chip using the embedded vector processing unit.
    On the other hand, the high-speed PlayPen setup represents a physical embodiment and environment that facilitates a large variety of experimental tasks on accelerated time scales.
    These experiments demonstrate the capabilites of the BrainScaleS-2 architecture for studying embodied cognition and provide many of the necessary tools required for this endeavor.
    \\
    \vspace{5pt}
    $\to$ KS: ACK.}

We outlined three agent-environment interaction scenarios using spiking neural networks emulated on the BrainScaleS-2 architecture.
In two of these, the sensors and actuators of an embodied neural agent, along with its environment, were also emulated on-chip using the embedded vector processing unit.
On the other hand, the high-speed PlayPen setup represents a physical embodiment and environment that facilitates a large variety of experimental tasks on accelerated time scales.
These experiments demonstrate the capabilites of the BrainScaleS-2 architecture for studying embodied cognition and provide many of the necessary tools required for this endeavor.

%% file: content/ack.tex
The author wishes to thank all present and former members of the Electronic Vision(s) research group contributing to the BSS-1 and BSS-2 hardware as well as software. 
We gratefully acknowledge funding from the European Union under grant agreements 604102, 720270, 785907 (HBP) and the Manfred St{\"a}rk Foundation.